\documentclass[showkeys,aps,amsmath,amssymb,twocolumn,groupedaddress,longbibliography,floats,final,postprint,superscriptaddress]{revtex4-1}
\usepackage{enumerate}
\usepackage{graphicx}
\usepackage{amsmath,amssymb,amsthm,enumitem}
\newcommand{\doublerightharpoonup}{%
  \rightharpoonup\mkern-10mu\rightharpoonup%
}
\usepackage{color}
\definecolor{LinkColor}{RGB}{199,21,133}
\usepackage[colorlinks=true,linkcolor=blue,anchorcolor=blue,citecolor=blue,urlcolor=blue]{hyperref}
\usepackage[polutonikogreek,english]{babel}

\newcommand{\headline}[1]{\textit{\textcolor{magenta}{#1.}}---}

\usepackage{hyperref}

\usepackage{epstopdf}
\usepackage{type1cm}
\usepackage{multirow}
\usepackage{times}

\def\scro{\mathcal{O}}

\begin{document}

\title{Extraordinary-log surface phase transition in the three-dimensional $XY$ model}
\author{Minghui Hu}
\affiliation{Department of Physics and Anhui Key Laboratory of Optoelectric Materials Science and Technology, Key Laboratory of Functional Molecular Solids, Ministry of Education, Anhui Normal University, Wuhu, Anhui 241000, China}
\author{Youjin Deng}
\email{yjdeng@ustc.edu.cn}
\affiliation{Hefei National Laboratory for Physical Sciences at Microscale and Department of Modern Physics,
University of Science and Technology of China, Hefei, Anhui 230026, China}
\affiliation{MinJiang Collaborative Center for Theoretical Physics,
College of Physics and Electronic Information Engineering, Minjiang University, Fuzhou 350108, China}
\author{Jian-Ping Lv}
\email{jplv2014@ahnu.edu.cn}
\affiliation{Department of Physics and Anhui Key Laboratory of Optoelectric Materials Science and Technology, Key Laboratory of Functional Molecular Solids, Ministry of Education, Anhui Normal University, Wuhu, Anhui 241000, China}

\date{\today}

\begin{abstract}
Universality is a pillar of modern critical phenomena. The standard scenario is that the two-point correlation algebraically decreases with the distance $r$ as $g(r) \sim r^{2-d-\eta}$, with $d$ the spatial dimension and $\eta$ the anomalous dimension. Very recently, a logarithmic universality was proposed to describe the extraordinary surface transition of O($N$) system. In this logarithmic universality, $g(r)$ decays in a power of logarithmic distance as $g(r) \sim ({\rm ln}r)^{-\hat{\eta}}$, dramatically different from the standard scenario. We explore the three-dimensional $XY$ model by Monte Carlo simulations, and provide strong evidence for the emergence of logarithmic universality. Moreover, we propose that the finite-size scaling of $g(r,L)$ has a two-distance behavior: simultaneously containing a large-distance plateau whose height decays logarithmically with $L$ as $g(L) \sim ({\rm ln}L)^{-\hat{\eta}'}$ as well as the $r$-dependent term $g(r) \sim ({\rm ln}r)^{-\hat{\eta}}$, with ${\hat{\eta}'} \approx {\hat{\eta}}-1$. The critical exponent $\hat{\eta}'$, characterizing the height of the plateau, obeys the scaling relation $\hat{\eta}'=(N-1)/(2\pi \alpha)$ with the RG parameter $\alpha$ of helicity modulus. Our picture can also explain the recent numerical results of a Heisenberg system. The advances on logarithmic universality significantly expand our understanding of critical universality.
\end{abstract}


\maketitle
\headline{Introduction}
Continuous phase transitions are ubiquitous, from the magnetic and superconducting transitions in real materials to the cooling of early universe. Near a second-order transition, a diverging correlation length emerges, and several macroscopic properties become independent of microscopic details of the system~\cite{stanley1999scaling,sachdev2007quantum,fernandez2013random}. Systems can be classified into few universality classes, depending on a small number of global features like symmetry, dimensionality and the range of interactions. Typically, physical quantities exhibit power-law behaviors governed by critical
exponents characteristic of a universality class. In particular, at criticality, the two-point correlation function $g(r)$ decays algebraically with the spatial distance $r$ as
\begin{equation}\label{tp0}
g(r) \sim r^{2-d-\eta},
\end{equation}
where $d$ is the spatial dimension and $\eta$ is the anomalous dimension. Power-law universality has been extensively verified and recognized as the standard scenario of critical phenomena~\cite{sachdev2007quantum,fernandez2013random,goldman2013percolation,svistunov2015superfluid}. Very recently, a novel logarithmic universality of criticality, drastically different from that encoded in (\ref{tp0}), was proposed in the context of surface critical behavior (SCB)~\cite{metlitski2020boundary}.

\begin{figure}
\includegraphics[width=7cm,height=4cm]{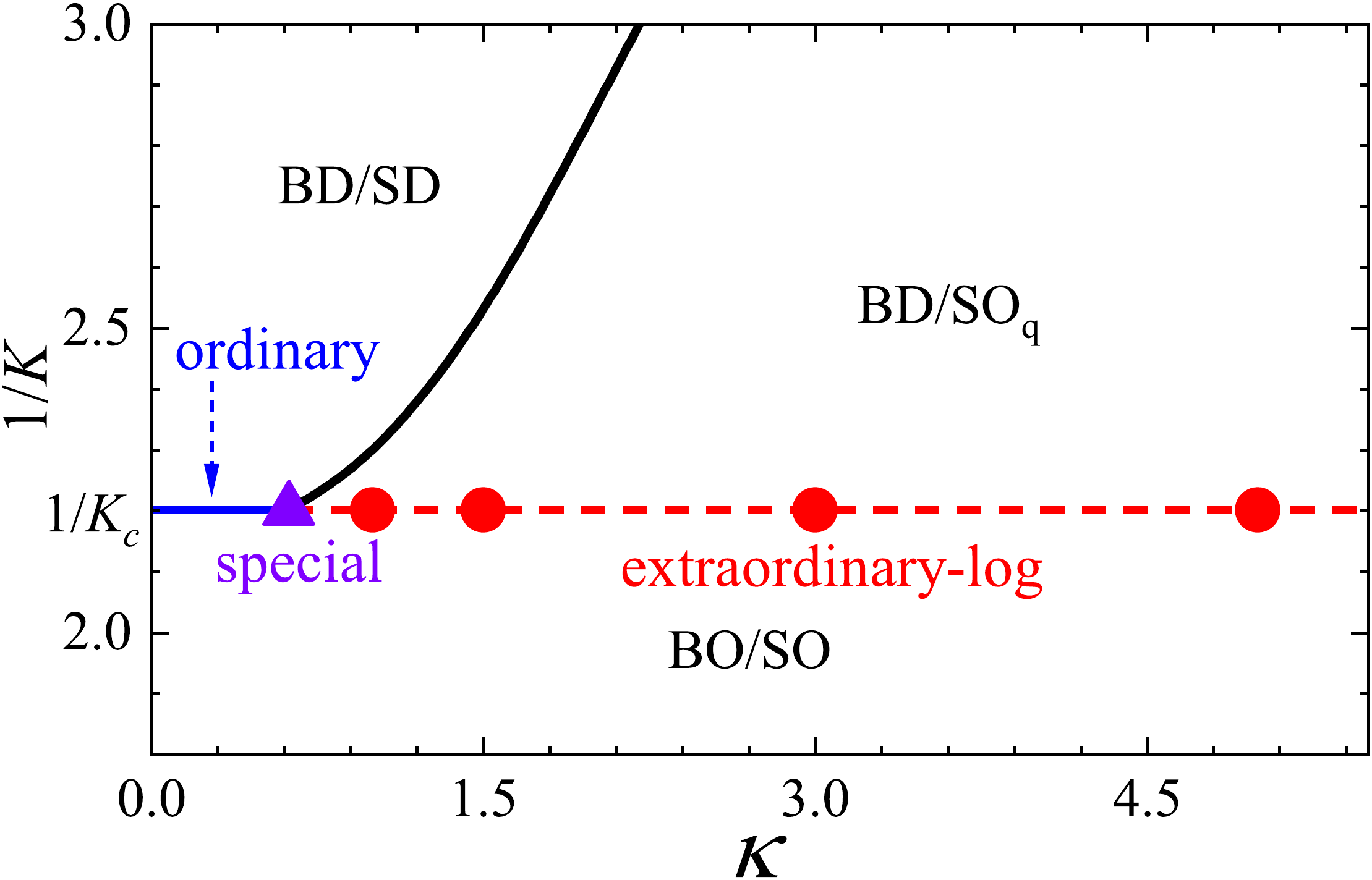}
\caption{Phase diagram of the $XY$ model~(\ref{Ham1}). The horizontal axis is for the surface coupling enhancement $\kappa$ and the vertical axis relates to the bulk coupling $K$ by $1/K$. Phases are denoted by the abbreviations BD (bulk disorder), BO (bulk order), SD (surface disorder), ${\rm SO_q}$ (surface quasi-long-range order), and SO (surface order). The ordinary, the extraordinary-log, and the SD-${\rm SO_q}$ critical lines meet together at the special critical point. The topology of the phase diagram is well known, but the nature of the extraordinary transition remains a puzzle~\cite{metlitski2020boundary}. Parameters denoted by red circles are used in this work to analyze the extraordinary-log universality class.}\label{fig_pd}
\end{figure}

\begin{figure*}
 \includegraphics[width=17cm,height=6cm]{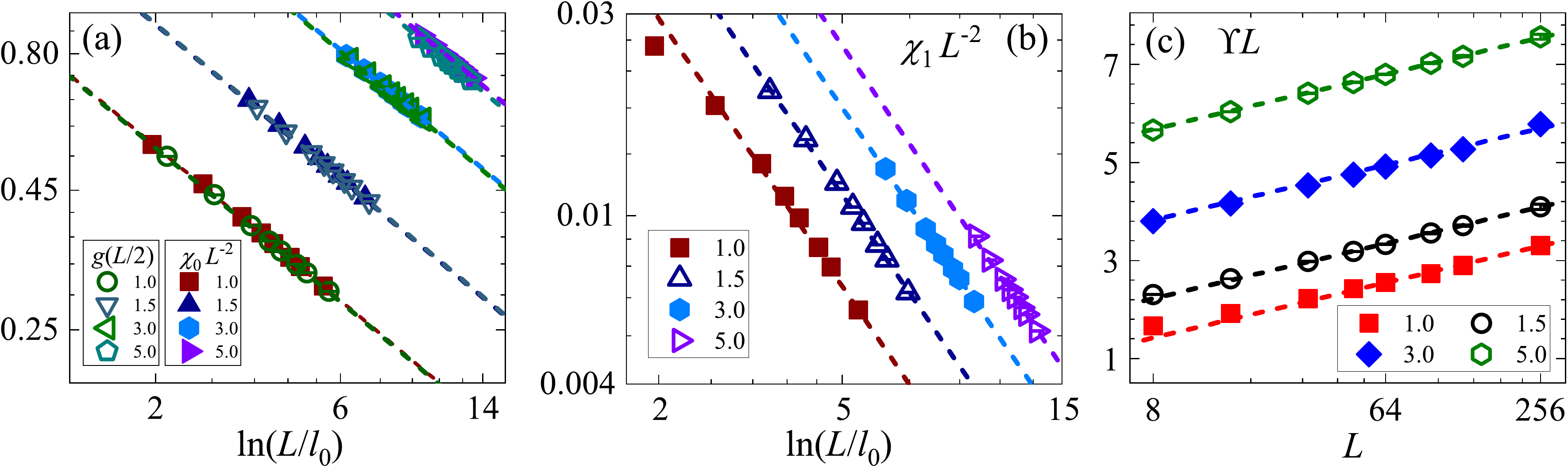}
\caption{Results for the extraordinary-log transitions at $\kappa=1, 1.5, 3$, and $5$. Statistical errors are much smaller than the size of symbols. (a) Log-log plot of the two-point correlation $g(L/2)$ and the scaled susceptibility $\chi_0 L^{-2}$ versus ${\rm ln} (L/l_0)$. The parameter $l_0$ is $\kappa$-dependent and obtained from least-squares fits. Dashed lines have the slope $-0.59$ and denote the critical exponent $\hat{\eta}'=0.59(2)$. (b) Log-log plot of the scaled magnetic fluctuations $\chi_1 L^{-2}$ versus ${\rm ln} (L/l_0)$. Dashed lines have the slope $-1.59$ and denote the exponent $\hat{\eta} \approx 1.59$. (c) Scaled helicity modulus $\Upsilon L$ versus $L$. The horizontal axis is in a log scale. Dashed lines have the slope $0.54$ and relate to the universal RG parameter $\alpha = 0.27(2)$ by $2\alpha$.}\label{fig_ex}
 \end{figure*}

SCB refers to the critical phenomenon occurring on the boundary of a critical bulk~\cite{binder1974surface,Ohno1984,landau1989monte,diehl1997theory,pleimling2004critical,deng2005surface,deng2006bulk,Dubail2009,zhang2017unconventional,ding2018engineering,Weber2018,metlitski2020boundary,Weber2021,toldin2020boundary}. Recent activities on SCB were partly triggered by the exotic surface effects of symmetry protected topological
phases~\cite{grover2012quantum,parker2018topological}. The O($N$) model exhibits rich SCBs including the {\it special}, {\it ordinary}, and {\it extraordinary} transitions, depending on $N$ and $d$~\cite{binder1974surface,Ohno1984,landau1989monte,diehl1997theory,pleimling2004critical,deng2005surface,deng2006bulk,Dubail2009,zhang2017unconventional,ding2018engineering,Weber2018,metlitski2020boundary,Weber2021,toldin2020boundary}. The situations at $d=3$ are extremely subtle and controversial~\cite{deng2005surface,deng2006bulk,metlitski2020boundary,Weber2021,toldin2020boundary}. Logarithmic universality of extraordinary transition was proposed for the three-dimensional O($N$) model with $2 \leq N < N_c$ by means of renormalization group (RG)~\cite{metlitski2020boundary}, whereas $N_c$ is not exactly known. It was predicted that the two-point correlation on surface decays logarithmically with $r$ as~\cite{metlitski2020boundary}
\begin{equation}\label{tp1}
g(r) \sim [{\rm ln}(r/r_0)]^{-\hat{\eta}},
\end{equation}
where $r_0$ is a non-universal constant. If $N$ is specified, the critical exponent
$\hat{\eta}$ is universal in extraordinary regime. The asymptotic form (\ref{tp1}) obviously differs from the standard scenario (\ref{tp0}). A quantum Monte Carlo study was performed for the SCB of a (2+1)-dimensional O(3) system~\cite{Weber2021}. However, both the logarithmic and the
extraordinary-power behavior~\cite{metlitski2020boundary} were not completely confirmed. By
contrast, compelling evidence for the logarithmic
behavior was obtained from a classical O(3) $\phi^4$ model~\cite{toldin2020boundary}.

In this work, we explore the extraordinary transition with $N=2$, which is the lower-marginal candidate for the logarithmic universality. We consider an extensive domain of extraordinary critical line, in which the universality of logarithmic behavior is confirmed. Moreover, we give a two-distance scenario for the finite-size scaling (FSS) of $g(r)$, where an $r$-independent plateau emerges at large distance. The height of the plateau exhibits a logarithmic FSS with the exponent $\hat{\eta}'$, which relates to the exponent $\hat{\eta}$ of $r$-dependent behavior by $\hat{\eta}' \approx \hat{\eta}-1$.

\headline{Main results}
We study the $XY$ model on simple-cubic lattices with the Hamiltonian~\cite{landau1989monte,deng2005surface}
\begin{equation}
{\cal H}/(k_{\rm B}T) = - {\sum_{\langle {\bf r}{\bf r'} \rangle}} K_{{\bf r}{\bf r'}}
\vec{S}_{\bf r} \cdot \vec{S}_{\bf r'} \; ,
\label{Ham1}
\end{equation}
where $\vec{S}_{\bf r}$ represents the $XY$ spin on site ${\bf r}$ and $K_{{\bf r}{\bf r'}}$ denotes the strength of nearest-neighbor ferromagnetic coupling. We impose open boundary conditions in one direction and periodic boundary conditions in other directions, hence a pair of open surfaces are specified. We set $K_{{\bf r}{\bf r'}}=K'$ if ${\bf r}$ and ${\bf r'}$ are on the same surface and $K_{{\bf r}{\bf r'}}=K$ otherwise. The surface coupling enhancement $\kappa$ is defined by $\kappa = (K'-K)/K$.

Figure~\ref{fig_pd} shows the phase diagram of model (\ref{Ham1}), which contains a
long-range-ordered surface phase in presence of ordered bulk, as well as disordered and critical quasi-long-range-ordered
surface phases in presence of disordered bulk. The critical lines meet together at the special transition
point. A characteristic feature for $N=2$ is the existence of the quasi-long-range-ordered phase, which is absent in $N=1$ and $N \ge 3$ situations.

Consider the quasi-long-range-ordered regime. As the bulk critical point $K_c$ is approached, namely $K \rightarrow K_c^-$, divergent bulk correlations emerge. A possible scenario is that the surface long-range order develops
at $K_c$ as a result of the effective interactions mediated by
long-range bulk correlations. This scenario can not be precluded
by the Mermin-Wagner theorem as the effective interactions
could be long-ranged. A previous study revealed~\cite{deng2005surface}
that the Monte Carlo data restricting to $L \leq 95$ ($L$ is linear size)
are not sufficient to preclude either discontinuous or continuous
surface transition across the extraordinary critical line; the former
implies long-range surface order at $K_c$.

By Monte Carlo sampling of the surface two-point correlation function $g(r) = \langle \vec{S}_0 \cdot \vec{S}_r \rangle$, we confirm the emergence of logarithmic universality in model~(\ref{Ham1}). As shown in Fig.~\ref{fig_ex}(a), the $L$ dependence of $g(L/2)$ obeys the scaling formula $g(L/2) \sim [{\rm ln}(L/l_0)]^{-\hat{\eta}'}$ with $\hat{\eta}'=0.59(2)$.

We analyze the surface magnetic fluctuations $\Gamma({\bf k}) = L^2 \langle ||\overset{\doublerightharpoonup}{m}({\bf k})||^2 \rangle$ with $\overset{\doublerightharpoonup}{m}({\bf k}) = (1/L^2)\sum_{\bf r} \vec{S}_{\bf r} e^{i{\bf k}\cdot{\bf r}}$, where the summation runs over sites on surface and ${\bf k}$ denotes a Fourier mode. As shown in Figs.~\ref{fig_ex}(a) and (b), the magnetic fluctuations $\chi_0 =\Gamma(0, 0)$ (susceptibility) and $\chi_1=\Gamma(2\pi/L, 0)$ have the distinct FSS behaviors $\chi_0 \sim L^2 [{\rm ln}(L/l_0)]^{-\hat{\eta}'}$ and $\chi_1  \sim L^2 [{\rm ln}(L/l_0)]^{-\hat{\eta}}$, with ${\hat{\eta}} \approx {\hat{\eta}'}+1$. Motivated by these observations as well as the two-distance scenarios in high-dimensional O($N$) critical systems~\cite{papathanakos2006finite,grimm2017geometric,zhou2018random,lv2020,fang2021Logarithmic} and quantum deconfined criticality~\cite{shao2016quantum}, we conjecture that the FSS of critical two-point correlation behaves as
\begin{equation}\label{tp}
g(r) \sim \begin{cases} [{\rm ln}(r/r_0)]^{-\hat{\eta}}, & {\rm ln}r \le  \scro [({\rm ln}L)^{\hat{\eta}'/\hat{\eta}}],  \\
[{\rm ln}(L/l_0)]^{-\hat{\eta}'},  & {\rm ln}r \ge   \scro [({\rm ln}L)^{\hat{\eta}'/\hat{\eta}}],\;
\end{cases}
\end{equation}
where $r_0$ and $l_0$ are non-universal constants. By (\ref{tp}), we point out two coexisting features:
the $r$-dependent behavior $[{\rm ln}(r/r_0)]^{-\hat{\eta}}$ and the large-distance $r$-independent
plateau $[{\rm ln}(L/l_0)]^{-\hat{\eta}'}$. Equation~(\ref{tp}) is an explanation for our numerical results and compatible with the FSS of second-moment correlation length at the extraordinary transition of O(3) model~\cite{toldin2020boundary,metlitski2021Private}. Recently, a two-distance scenario was used to describe the two-point correlation of O($n$) model at a marginal situation (the upper critical dimensionality)~\cite{lv2020} and confirmed by large-scale simulations on hyper-cubic lattices up to $768^{4}$ sites~\cite{fang2021Logarithmic}. The open surfaces of model~(\ref{Ham1}) are at the lower critical dimensionality ($d_s=2$) and also belong to marginal situations.

\begin{figure}
 \includegraphics[width=8cm,height=6cm]{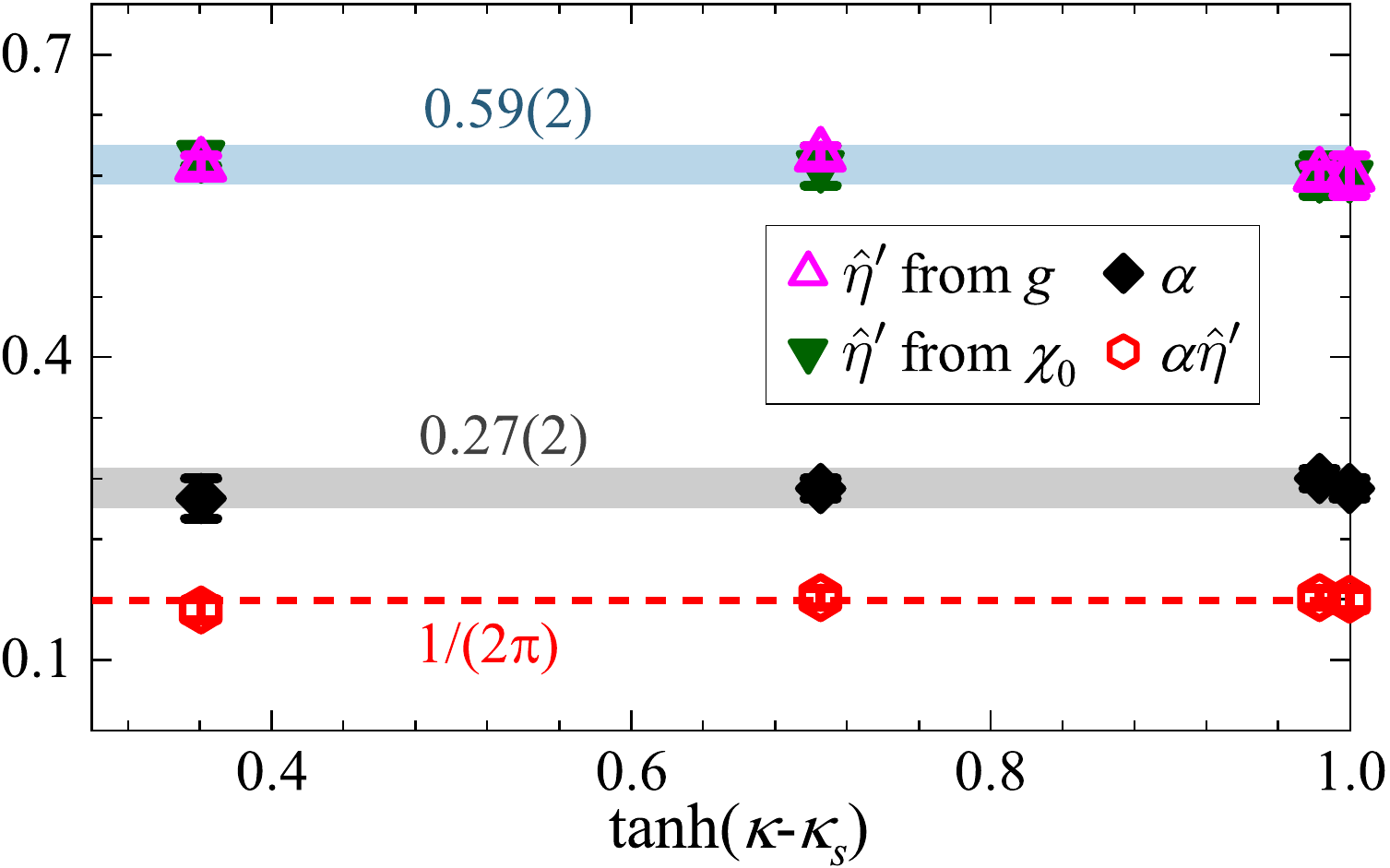}
\caption{The critical exponent $\hat{\eta}'$ estimated from $g(L/2)$ and $\chi_0$, the RG parameter $\alpha$ from $\Upsilon$, and their product $\alpha \hat{\eta}'$. Error bars are plotted with symbols. The shadowed areas, whose heights represent two error bars, denote the ranges of our final estimates for $\hat{\eta}'$ and $\alpha$. The red dashed line denotes the predicted value $\alpha \hat{\eta}'=1/(2\pi)$ by RG.}\label{fig_ep}
 \end{figure}

We confirm the scaling relation between $\hat{\eta}'$ and the RG parameter of helicity modulus. The helicity modulus $\Upsilon$ measures the response of a system to a twist in boundary conditions~\cite{fisher1973}. The definition is given in the Supplementary Materials (SM). Figure~\ref{fig_ex}(c) demonstrates that $\Upsilon$ scales as $\Upsilon L \sim 2 \alpha {\rm ln}L$ with the RG parameter $\alpha=0.27(2)$. Figure~\ref{fig_ep} simultaneously illustrates the universality of $\hat{\eta}'$ and $\alpha$ in the extraordinary regime. Meanwhile, the scaling relation $\alpha \hat{\eta}'=1/(2 \pi)$ is evidenced, conforming to the predicted form~\cite{metlitski2020boundary}
\begin{equation}\label{qp}
\hat{\eta}'=\frac{N-1}{2 \pi \alpha}.
\end{equation}
According to (\ref{tp}), the exponent $\hat{\eta}'$ characterizes the FSS of the height of the plateau. Equation~(\ref{qp}) is not exactly the original prediction in Ref.~\cite{metlitski2020boundary}, where the exponent $\hat{\eta}$ for $r$-dependent behavior obeys the relation $\hat{\eta}=(N-1)/(2 \pi \alpha)$.

\headline{Technique aspects}
To explore the SCB, we fix the bulk coupling strength at $K_c$. Previously, two of us and coworkers performed simulations utilizing the Prokof'ev-Svistunov worm algorithms~\cite{prokof2001worm,prokof2010worm} on periodic simple-cubic lattices with $L_{\rm max}=512$, and obtained $1/K_c=2.201\,844\,1(5)$~\cite{Xu2019}. This estimate was confirmed by an independent Monte Carlo study~\cite{hasenbusch2019monte}. Here, we simulate model (\ref{Ham1}) at $1/K_c=2.201\,844\,1$ using Wolff's cluster algorithm~\cite{Wolff1989} on simple-cubic lattices with $L_{\rm max}=256$. The original procedure in Ref.~\cite{Wolff1989} is adapted to model~(\ref{Ham1}). We analyze the extraordinary transitions at $\kappa=1$, $1.5$, $3$, and $5$, and the special transition at $\kappa_s=0.622\,2$~\cite{deng2005surface}. For each $\kappa$, the number of Wolff updating steps is up to $1.2 \times 10^8$ for $L \leq 32$ and ranges from $1.7 \times 10^8$ to $6.1 \times 10^8$ for $L\geq 48$. See SM for details, which includes Refs.~\cite{JANKE1990306,Krauth}.

Our conclusions are based on FSS analyses performed by using least-squares fits. Following Refs.~\cite{hasenbusch2019monte,hasenbusch2020monte}, the function $curve\_fit()$ in $Scipy\_library$ is adopted. For caution, we compare the fits with the benchmarks from implementing Mathematica's $NonlinearModelFit$ function as Ref.~\cite{salas2020phase}. The fits with the Chi squared per degree of freedom $\chi^2/{\rm DF} \sim 1$ are preferred. We do not trust any single fit and final conclusions are drawn based on comparing the fits that are stable against varying $L_{\rm min}$, the minimum size incorporated.

\begin{table}
\begin{center}
\caption{Estimates of the critical exponent $\hat{\eta}'$ and the RG parameter $\alpha$ for the extraordinary-log transition at $\kappa=1$. $\hat{\eta}'$ is estimated from the scaling formulae $g(L/2) \sim [{\rm ln}(L/l_0)]^{-\hat{\eta}'}$ and $\chi_0 \sim L^2 [{\rm ln}(L/l_0)]^{-\hat{\eta}'}$, and $\alpha$ is determined from $\Upsilon L =2 \alpha {\rm ln}L+A+BL^{-1}$.}
				\label{fit_result_1}
				\begin{tabular}[t]{p{1.4cm}p{1.2cm}p{1.5cm}p{1.7cm}p{0.9cm}}
					\hline
					\hline
					 &$L_{\rm min}$               & $\chi^2$/DF &    $\hat{\eta}'$ or $\alpha$   &   $l_0$ or $A$ \\
					\hline
					$g(L/2)$ & 16     & 2.91/4                     & 0.596(2) & 0.94(1)\\
					& 32     & 0.66/3                   & 0.592(3) &  0.97(2)\\
					& 48     & 0.58/2                     & 0.591(5) & 0.98(4)\\
		
					\hline
				$\chi_0$	& 32     & 3.46/3                    & 0.603(2) &  1.13(2)\\
					& 48     & 0.08/2                   & 0.598(4) & 1.18(3)\\
					& 64     & 0.02/1                    & 0.597(5) & 1.19(5)\\
				
					\hline
				$\Upsilon$	& 8     & 5.46/4                    & 0.255(3) &0.41(2) \\
					& 16    & 3.33/3                    & 0.265(7) &0.32(6)\\
					& 32    & 2.51/2                    & 0.25(2) & 0.4(2)\\
	
					\hline
					\hline
				\end{tabular}
\end{center}
\end{table}

\begin{figure}
 \includegraphics[width=8cm,height=8cm]{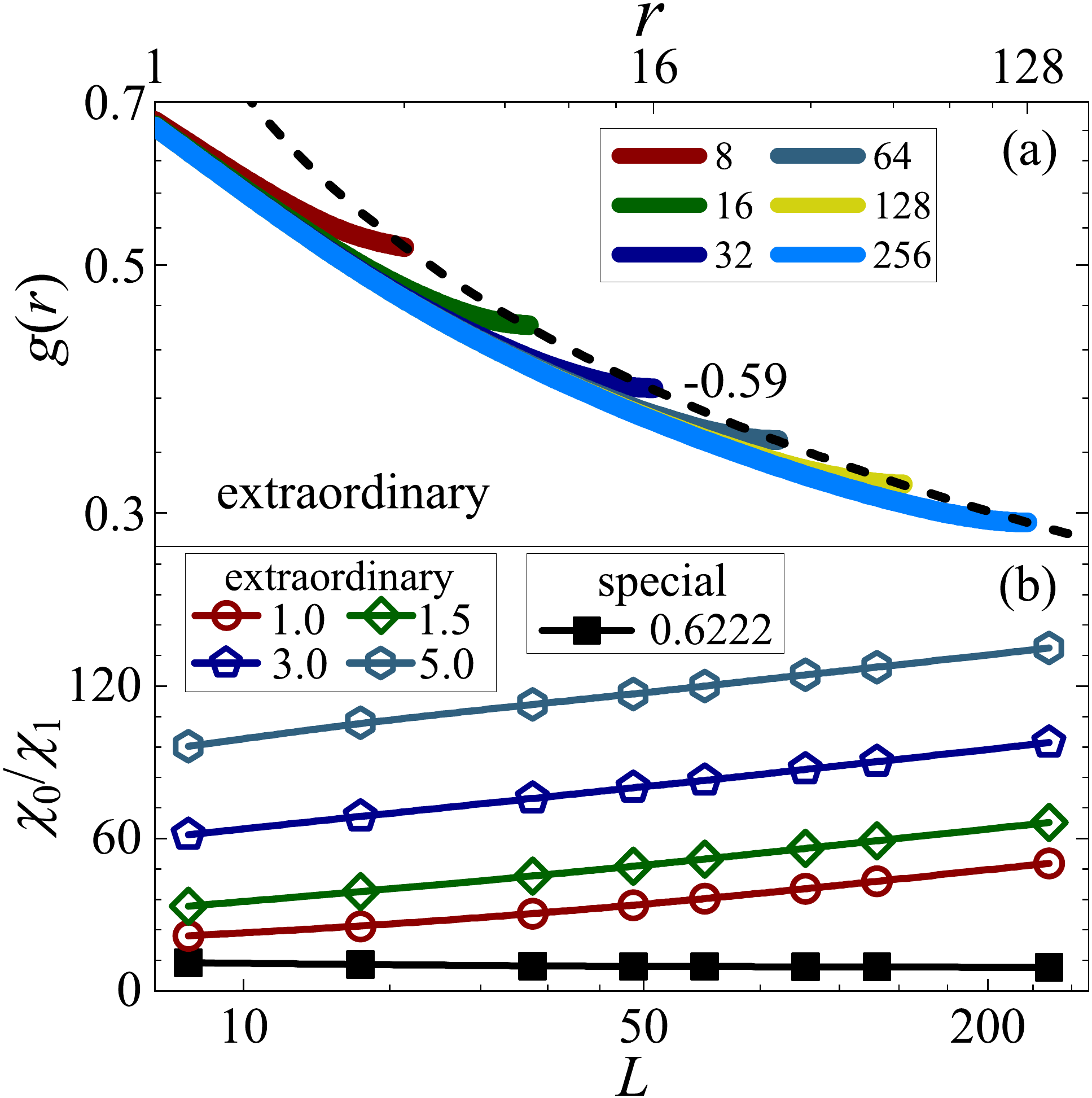}
\caption{(a) The two-point correlation $g(r)$ for the extraordinary-log transition at $\kappa=1$ with $L=8, 16, 32, 64, 128$, and $256$. The dashed line denotes the logarithmic decaying $[{\rm ln}(L/l_0)]^{-0.59}$ in the large-distance limit. (b) The ratio $\chi_0/\chi_1$ of magnetic fluctuations versus $L$ for the extraordinary-log transitions at $\kappa=1$, $1.5$, $3$, and $5$, and for the special transition at $\kappa_s=0.622\,2$. In both panels, statistical errors are much smaller than the sizes of the data points.}\label{fig4}
\end{figure}

\headline{Emergence of logarithmic universality}
Figure~\ref{fig4}(a) demonstrates the two-point correlation function
$g(r)$ for the extraordinary transition at $\kappa=1$. The large-distance behavior can be monitored by the $L$ dependence
of $g(L/2)$. According to Eq.~(\ref{tp}), we have a
scaling formula $g(L/2) \sim [{\rm ln}(L/l_0)]^{-\hat{\eta}'}$. We perform least-squares fits to this formula and obtain
$\hat{\eta}'= 0.596(2)$, $l_0=0.94(1)$, and $\chi^2/{\rm DF} \approx 0.73$, with
$L_{\rm min}=16$. As $L_{\rm min}$ is varied, preferred fits are also obtained (Table~\ref{fit_result_1}). By comparing the fits, our final estimate of $\hat{\eta}'$ for $\kappa=1$ is $\hat{\eta}'=0.59(1)$. In the SM, we present similar
analyses for $\kappa= 1.5, 3$ and $5$, for which the final estimates are $\hat{\eta}'=0.60(1)$ ($\kappa=1.5$), $0.58(1)$ ($\kappa=3$) and $0.58(2)$ ($\kappa=5$). It is therefore confirmed that $g(L/2)$ obeys the
logarithmic scaling $g(L/2) \sim [{\rm ln}(L/l_0)]^{-\hat{\eta}'}$, with a universal exponent $\hat{\eta}'=0.59(2)$. As displayed in the SM, the fits by the conventional power-law ansatz (\ref{tp0}) have poor qualities and give unstable results.

\headline{Existence of two distinct exponents}
For a verification of Eq.~(\ref{tp}), we analyze the FSS of surface magnetic fluctuations. In the Monte Carlo simulations, we sample $\chi_2=\Gamma(2\pi/L, 2\pi/L)$ as well as $\chi_0$ and $\chi_1$.

According to~(\ref{tp}), an $r$-independent plateau emerges at large distance. This plateau contributes to the magnetic fluctuations at zero mode but not to those at non-zero modes. The ratio $\chi_0/\chi_1$ at extraordinary transitions is shown in Fig.~\ref{fig4}(b). As $L \rightarrow \infty$, the ratio keeps increasing, implying distinct FSS of $\chi_0$ and $\chi_1$.

More precisely, $\chi_0$ is expected to scale as $\chi_0 \sim L^2[{\rm ln}(L/l_0)]^{-\hat{\eta}'}$.
The results of scaling analyses for $\kappa=1$ are illustrated
in Table~\ref{fit_result_1} and those for $\kappa= 1.5$, $3$, and $5$ are given in
SM. Comparing preferred fits, we obtain $\hat{\eta}' = 0.60(1)$ ($\kappa=1$),
$0.59(2)$ ($\kappa=1.5$), $0.58(2)$ ($\kappa=3$), and $0.58(1)$ ($\kappa=5$). These estimates of $\hat{\eta}'$ agree well with those determined from the $L$ dependence of $g(L/2)$, hence the final result $\hat{\eta}'=0.59(2)$ is confirmed.

We analyze the magnetic fluctuations $\chi_1$ and $\chi_2$ at nonzero Fourier modes by performing fits to $\chi_{{\bf k} \neq 0} \sim L^2 [{\rm ln}(L/l_0)]^{-\hat{\eta}}$. We confirm the drastic decays of $\chi_1 L^{-2}$ and $\chi_2 L^{-2}$ upon increasing ${\rm ln}L$. For reducing the uncertainties of fits, we fix $l_0$ at those obtained from the scaling analyses of $\chi_0$, and estimate $\hat{\eta} \approx 1.7$ over $\kappa=1, 1.5, 3$, and $5$. From the log-log plot of $\chi_1 L^{-2}$ versus ${\rm ln}(L/l_0)$ in Fig.~\ref{fig_ex}(b), it is seen that the data nearly scale as $\chi_1 L^{-2} \sim [{\rm ln}(L/l_0)]^{-\hat{\eta}}$ with $\hat{\eta} \approx 1.59$. Similar result is obtained for $\chi_2 L^{-2}$ (SM). Hence, $\chi_1$ and $\chi_2$ obey the logarithmic FSS formula $\chi_{{\bf k} \neq 0} \sim L^2 [{\rm ln}(L/l_0)]^{-\hat{\eta}}$, with $\hat{\eta} \approx 1.6$.

Our results for the FSS of $\chi_0$ and $\chi_1$ are also compatible with the Monte Carlo data~\cite{toldin2020boundary,metlitski2021Private} of the second-moment correlation length $\xi_{\rm 2nd}$, which scales as $(\xi_{\rm 2nd}/L)^2 \sim (\chi_0/\chi_1-1) \sim {\rm ln}L$. The relation $\hat{\eta}=\hat{\eta}'+1$ is implied.

As $\hat{\eta}$ is much larger than $\hat{\eta}'$, the two-distance scenario~(\ref{tp}) indicates that the $r$-dependent contribution decays fast. It explains the profile of $g(r)$ in Fig.~\ref{fig4}(a), where the large-distance plateau dominates.

By contrast, the special transition at $\kappa_s$ belongs to the standard scenario (\ref{tp0}) of continuous transition. The $r$-dependent behavior converges to the power law $g(r)\sim r^{-\eta}$, which is comparable with the contribution from $g(L/2) \sim L^{-\eta}$. Moreover, the magnetic renormalization exponent $y_h$ relates to the anomalous dimension $\eta$ by $y_h=(4-\eta)/2$, and the magnetic fluctuations $\chi_0$, $\chi_1$, and $\chi_2$ {\it all} scale as $L^{2y_h-2}$. As shown in Fig.~\ref{fig4}(b), the ratio $\chi_0/\chi_1$ at $\kappa_s$ converges fast to a constant upon increasing $L$. More results for $g(r)$, $\chi_0$, $\chi_1$ and $\chi_2$ are given in SM.

\headline{Scaling relation}
It was predicted~\cite{metlitski2020boundary,metlitski2021Private} that the scaled helicity modulus $\Upsilon L$ diverges logarithmically as $\Upsilon L  \sim 2 \alpha {\rm ln}L$, with $\alpha$ a universal RG parameter. Further, the universal form (\ref{qp}) of scaling relation was established~\cite{metlitski2020boundary}. The form is supported by the Monte Carlo results of an O(3) $\phi^4$ model~\cite{toldin2020boundary}.

We sample $\Upsilon$ of model~(\ref{Ham1}) by Monte Carlo simulations. The dependence of $\Upsilon L$ on ${\rm ln}L$ is shown in Fig.~\ref{fig_ex}(c) for $\kappa=1, 1.5, 3$, and $5$. For each $\kappa$, a nearly linear dependence is observed in large-$L$ regime. Further, we perform a FSS analysis of $\Upsilon$ according to $\Upsilon L =2 \alpha {\rm ln}L+A+BL^{-1}$, where $A$ and $B$
are constants. We explore the situations with and without the
correction term $BL^{-1}$ separately. Stable fits are achieved, with the final estimates of $\alpha$ being $\alpha = 0.26(2)$
($\kappa=1$), $0.27(1)$ ($\kappa=1.5$), $0.28(1)$ ($\kappa=3$), and $0.27(1)$
($\kappa=5$). Comparing these estimates, the universal value of $\alpha$ is determined to be $\alpha=0.27(2)$.

As shown in Fig.~\ref{fig_ep}, the scaling relation (\ref{qp}) between $\alpha$ and $\hat{\eta}'$ is confirmed. According to (\ref{tp}), $\hat{\eta}'$ characterizes the logarithmic FSS for the height of plateau.

\headline{Discussions}
We provide strong evidence for the emergence of the extraordinary-log universality class~\cite{metlitski2020boundary}. We propose the two-distance scenario (\ref{tp}) for the FSS of two-point correlation function, where a large-distance plateau emerges. The height of the plateau decays logarithmically with $L$ by the exponent $\hat{\eta}'$, which obeys the scaling relation (\ref{qp}) with the RG parameter of helicity modulus. The two-distance scenario is supported not only by the Monte Carlo data for $N=2$ of this work, but also by the results for $N=3$ in Ref.~\cite{toldin2020boundary}.

A variety of open questions arise. First, it is shown essentially that a two-dimensional $XY$ system with {\it finely tuned} long-range interactions exhibits logarithmic universality. Is it possible to formulate the interactions in a microscopic Hamiltonian? Second, is there a classical-quantum mapping for the two-distance scenario that holds at the O($N$) quantum critical points~\cite{ding2018engineering,Weber2018,Weber2021}? Third, as shown in Ref.~\cite{fang2021Logarithmic}, the introduction of unwrapped distance is crucial for verifying the short-distance behavior in two-distance scenario. The behavior of unwrapped distance in the extraordinary-log universality remains unclear. Finally, we note that, as recently observed for the five-dimensional Ising model~\cite{FangComplete}, lattice sites can be decomposed into clusters, and interesting geometric phenomena associated with the two-distance scenario may arise~\cite{Sun2021}.

\begin{acknowledgments}
We thank Max Metlitski for useful comments and for sharing an unpublished note~\cite{metlitski2021Private} with us. This work has been supported by the National Natural Science Foundation of China (under Grant Nos.~11774002, 11625522, and 11975024), the Science and Technology Committee of Shanghai (under grant No. 20DZ2210100), the National Key R\&D Program of China (under Grant No. 2018YFA0306501), and the Education Department of Anhui.
\end{acknowledgments}

\bibliography{papers}

\end{document}


\title{Supplementary Materials for: \\ ``Extraordinary-log surface phase transition in the three-dimensional $XY$ model"}

\author{Minghui Hu}
\affiliation{Department of Physics and Anhui Key Laboratory of Optoelectric Materials Science and Technology, Key Laboratory of Functional Molecular Solids, Ministry of Education, Anhui Normal University, Wuhu, Anhui 241000, China}
\author{Youjin Deng}
\email{yjdeng@ustc.edu.cn}
\affiliation{Hefei National Laboratory for Physical Sciences at Microscale and Department of Modern Physics,
University of Science and Technology of China, Hefei, Anhui 230026, China}
\affiliation{MinJiang Collaborative Center for Theoretical Physics,
College of Physics and Electronic Information Engineering, Minjiang University, Fuzhou 350108, China}
\author{Jian-Ping Lv}
\email{jplv2014@ahnu.edu.cn}
\affiliation{Department of Physics and Anhui Key Laboratory of Optoelectric Materials Science and Technology, Key Laboratory of Functional Molecular Solids, Ministry of Education, Anhui Normal University, Wuhu, Anhui 241000, China}

\date{\today}	
\maketitle		

In this Supplementary Materials (SM), we present more details for the Monte Carlo simulations as well as the finite-size scaling (FSS) of the extraordinary-log transition and the special transition.

The Monte Carlo data of this work are all obtained from
simulations using the Wolff's cluster algorithm~\cite{Wolff1989}. The bond between a pair of nearest-neighbor sites (${\bf r}$, ${\bf r'}$) is randomly occupied with the probability
\begin{equation}
p({\bf r}, {\bf r'})=\max [0, 1-\exp (-2 K'S^{(y)}_{\bf r}
 S^{(y)}_{\bf r'})]
\end{equation}
if ${\bf r}$ and ${\bf r'}$ are on the same surface, and with the probability
\begin{equation}
   p({\bf r}, {\bf r'})=\max [0, 1-\exp (-2 KS^{(y)}_{\bf r}
 S^{(y)}_{\bf r'})] \; .
\end{equation}
The spins in the Wolff cluster are flipped along the randomly chosen $y$ direction.
For $\kappa \in \{0.6222, 1, 1.5, 3, 5\}$,
the number of Wolff steps
is up to $1.2 \times 10^8$ for $L \leq 32$ and
ranges from $1.7 \times 10^8$ to $6.1 \times 10^8$ for $L=48, 64, 96, 128, 256$,
and the initial one sixth of the simulations are used for thermalization.

The procedure for estimating statistical errors is as follows. At the bulk criticality,
the average size of Wolff clusters is about $C \approx L^2$,
and, in our simulations, samples are taken for every $L$ Wolff steps,
which effectively correspond to a Monte Carlo sweep.
Since the autocorrelation times of the Wolff algorithm
are respectively $\tau_{\chi} \approx 1.7$ and $\tau_{e} \approx 1.3 L^{0.25}$
for susceptibility and energy~\cite{JANKE1990306},
the subsequent samples are nearly independent for the correlation function.
To deal with the residual correlation in the Markov chain,
we perform the widely-used data-bunching technique~\cite{Krauth},
which replaces adjacent samples by their average value.
The final error is then computed from a series of independent Markov chains
run at different CPUs with different random-number seeds.
With the estimated error for each Markov chain,
the residual $\chi^2$ are calculated to further check the quality of Markov chains.

In what follows, FSS analyses for the extraordinary-log transition and the special transition are given in Sec.~\ref{se1} and \ref{se2}, respectively.

\section{Extraordinary-log transition}\label{se1}
For exploring the extraordinary transition, we consider the surface coupling enhancements $\kappa=1, 1.5, 3$ and $5$.

\begin{figure*}
\includegraphics[height=17cm,width=12cm]{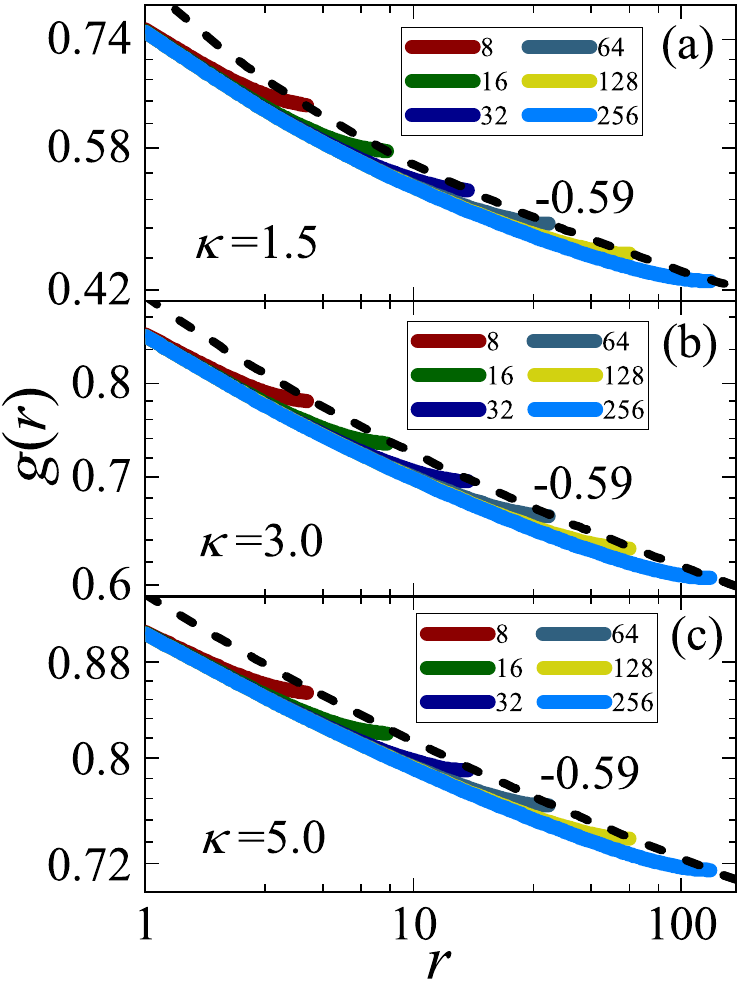}
\caption{The two-point correlation $g(r)$ for the extraordinary transitions at $\kappa=1.5$, $3$, and $5$ with $L=8, 16, 32, 64, 128$, and $256$. The dashed lines denote the logarithmic decaying $[{\rm ln}(L/l_0)]^{-0.59}$ in the large-distance limit. Statistical errors are much smaller than the width of solid lines.}~\label{SM_fig1}
\end{figure*}

The two-point correlation function $g(r)$ for $\kappa=1$ has been shown in Fig.~4(a) of the main text. In this SM, we present the results for $\kappa=1.5, 3$ and $5$ in Fig.~\ref{SM_fig1}, with $L_{\rm max}=256$. We fit the $g(L/2)$ data to the ansatz
\begin{equation}\label{fite1}	
g(L/2) = A[{\rm ln}(L/l_0)]^{-\hat{\eta}'},						
\end{equation}
where $\hat{\eta}'$ is a universal exponent. $A$ and $l_0$ are non-universal constants. The results for the fits are given in Table~\ref{fit_result_1A}. It is found, for each $\kappa$, that the fits with $\chi^2/{\rm DF} \sim 1$ are achieved. We evaluate the uncertainties of fits by incorporating both statistical and systematic errors. The latter is estimated by examining the stability of fits against varying $L_{\rm min}$. By comparing the estimates for considered enhancements, our final result of $\hat{\eta}'$ is $\hat{\eta}'=0.59(2)$.

The least-squares fits according to power-law hypothesis are also performed. First, we fit $g(L/2)$ to the power-law form $g(L/2)=AL^{-\eta}$. The results of the fits are given in Table~\ref{fit_result_1B}. For each $\kappa$, as $L_{\rm min}$ increases to $L_{\rm min}=96$, $\chi^2/{\rm DF}$
is still as large as $\chi^2/{\rm DF}>100$. Then, by considering corrections-to-scaling, we fit the data to the formula $g(L/2)=AL^{-\eta}(1+BL^{-\omega})$ and try various values for the correction exponent $\omega$. The results with $\omega=1$ and $1/2$ are given in Table~\ref{fit_result_1C}. For $\omega=1$, $\chi^2/{\rm DF}$ is generally large and the result of $\eta$
is unstable against increasing $L_{\rm min}$. For $\omega=1/2$, the fits with certain $L_{\rm min}$ can marginally satisfy the criterion $\chi^2/{\rm DF} \sim 1$. However, the results of $\eta$ for different $\kappa$ are not equal, violating the hypothesis
of a single universality class. It is also noted that the fitting value of exponent $\eta$ is very small. These observations provide strong evidence that the conventional power-law scaling is unlikely.

The FSS of the susceptibility $\chi_0$ is shown in Fig.~2(a) of the main text.
We analyze the critical susceptibility with the scaling formula
\begin{equation}\label{fite2}					
\chi_0 = A L^2[{\rm ln}(L/l_0)]^{-\hat{\eta}'}.		
\end{equation}
The details of fits are given in Table~\ref{fit_result_2}. The result $\hat{\eta}'=0.59(2)$ is confirmed.

\begin{figure*}
\includegraphics[height=10.5cm,width=12cm]{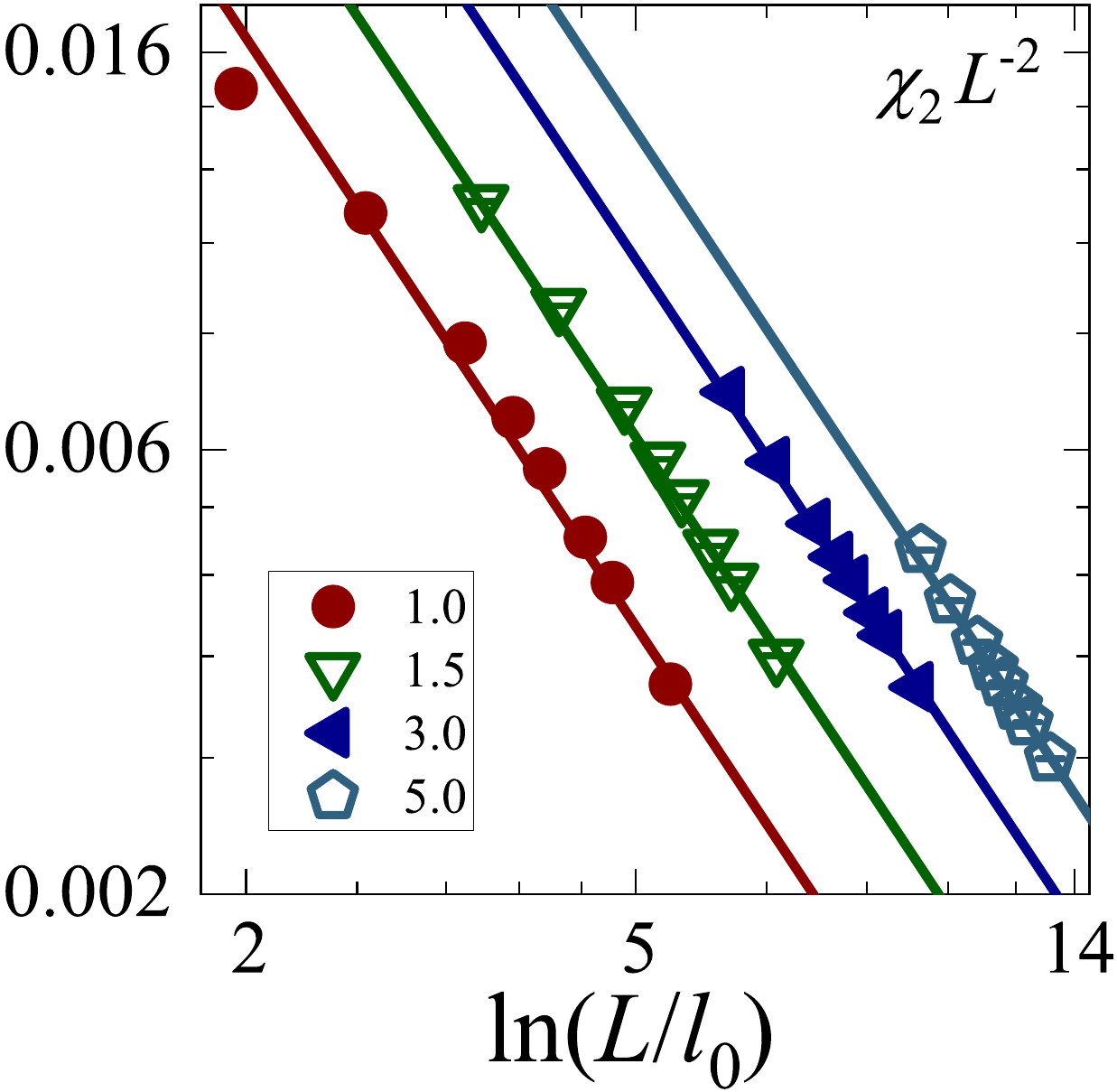}
\caption{Log-log plot of the scaled magnetic fluctuations $\chi_2 L^{-2}$ versus ${\rm ln} (L/l_0)$. Lines have the slope $-1.59$ and denote the fits to $\chi_2 L^{-2} \sim [{\rm ln}(L/l_0)]^{-\hat{\eta}}$ with $\hat{\eta} \approx 1.59$. Statistical errors are much smaller than the size of symbols.}~\label{SM_fig2}
\end{figure*}

For $\chi_1$ and $\chi_2$, we perform fits to the scaling formula
\begin{equation}\label{fite3}	
\chi_{{\bf k} \neq 0}=A  L^2 [{\rm ln}(L/l_0)]^{-\hat{\eta}}.
\end{equation}
The results are given in Table~\ref{fit_result_3a} and we have $\hat{\eta} > \hat{\eta}'$. It is practically difficult to obtain a precise estimate of $\hat{\eta}$. For reducing uncertainties, we fix $l_0$ at those produced by the preferred fits of $\chi_0$ to (\ref{fite2}). Accordingly, we find $\hat{\eta} \approx 1.7$ over $\kappa=1, 1.5, 3$ and $5$ (Table~\ref{fit_result_3b}). From the log-log plot of $\chi_1 L^{-2}$ versus ${\rm ln}(L/l_0)$ in Fig.~2(b) of the main text, we find that the data nearly scale as $\chi_1 L^{-2} \sim [{\rm ln}(L/l_0)]^{-\hat{\eta}}$ with $\hat{\eta} \approx 1.59$. As shown in Fig.~\ref{SM_fig2}, similar result is found for $\chi_2 L^{-2}$. In short, $\chi_1$ and $\chi_2$ obey the logarithmic scaling formula $\chi_{{\bf k} \neq 0} \sim L^2 [{\rm ln}(L/l_0)]^{-\hat{\eta}}$, with $\hat{\eta} \approx 1.6$.

Finally, we analyze the FSS of the helicity modulus $\Upsilon$, which is defined as~\cite{toldin2020boundary}
\begin{equation}
\Upsilon = \frac{1}{L^3}(\langle E \rangle-\langle T^2\rangle),
\end{equation}	
with
\begin{equation}
\begin{aligned}
E = &K\sum_{\bf{r}\in {\rm bulk}}\vec{S}_{\bf{r}}\cdot\vec{S}_{\bf{r}+\bf{e}_x} + K' \sum_{\bf{r}\in {\rm surfaces}}\vec{S}_{\bf{r}}\cdot\vec{S}_{\bf{r}+\bf{e}_x},\\	
T = &K\sum_{\bf{r}\in {\rm bulk}}(S_{\bf{r}}^a S_{\bf{r}+\bf{e}_x}^b - S_{\bf{r}}^b S_{\bf{r}+\bf{e}_x}^a)\\
&+ K' \sum_{\bf{r}\in {\rm surfaces}}(S_{\bf{r}}^a S_{\bf{r}+\bf{e}_x}^b -S_{\bf{r}}^b S_{\bf{r}+\bf{e}_x}^a).
\end{aligned}		
\end{equation}	
Here, $a$ and $b$ denote the two components of the two-dimensional spin vector. $\bf{e}_x$ denotes the unit vector along an edge direction of surfaces. The first summation in $E$ and $T$ are over sites in bulk and the second summations run over sites on surfaces. For each $\kappa$, we perform fits according to the ansatz
\begin{equation}\label{fite4}
\Upsilon L = 2\alpha{\rm ln}L+A+{BL^{-1}},							
\end{equation}
where $\alpha$ is universal and $A$ is non-universal. $BL^{-1}$ stands for finite-size corrections. The results of the fits are summarized in Table~\ref{fit_result_4}. Notice that the estimates of $\alpha$ for different $\kappa$ are compatible. By comparing the estimates, our final result is $\alpha=0.27(2)$.

\section{Special transition}\label{se2}

At the special transition, we analyze the $r$-dependent behavior of $g(r)$ as well as the FSS of $g(L/2), \chi_0, \chi_1$ and $\chi_2$. Notice that the transition point $\kappa_s=0.622\,2(3)$ and the magnetic renormalization exponent $y_h=1.675(1)$ have been given in literature~\citep{deng2005surface}. The latter relates to the anomalous dimension $\eta \approx 0.650$.

\begin{figure*}
\includegraphics[height=10cm,width=12cm]{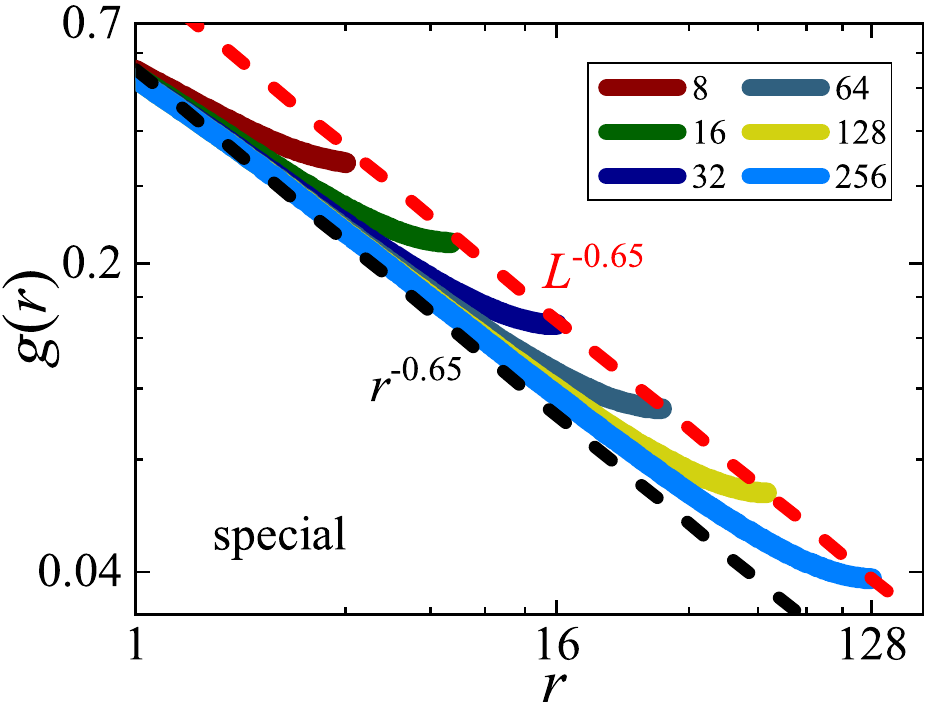}
\caption{Log-log plot of the two-point correlation $g(r)$ versus $r$ at the special transition point $\kappa_s=0.622\,2$. Statistical errors are much smaller than the width of solid lines.}~\label{SM_fig3}
\end{figure*}

\begin{figure*}
\includegraphics[height=10cm,width=12cm]{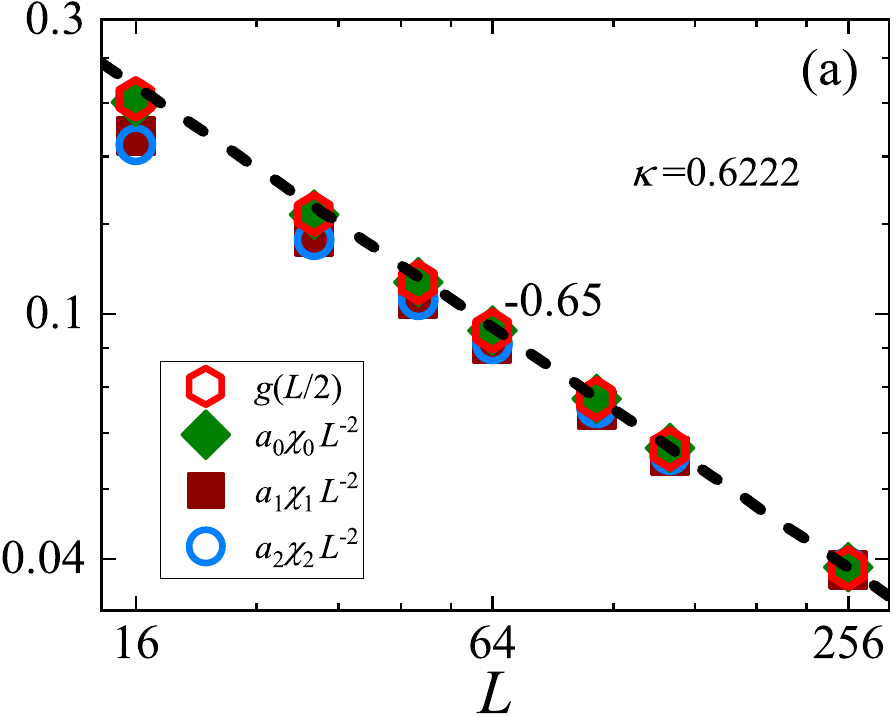}
\includegraphics[height=10cm,width=12cm]{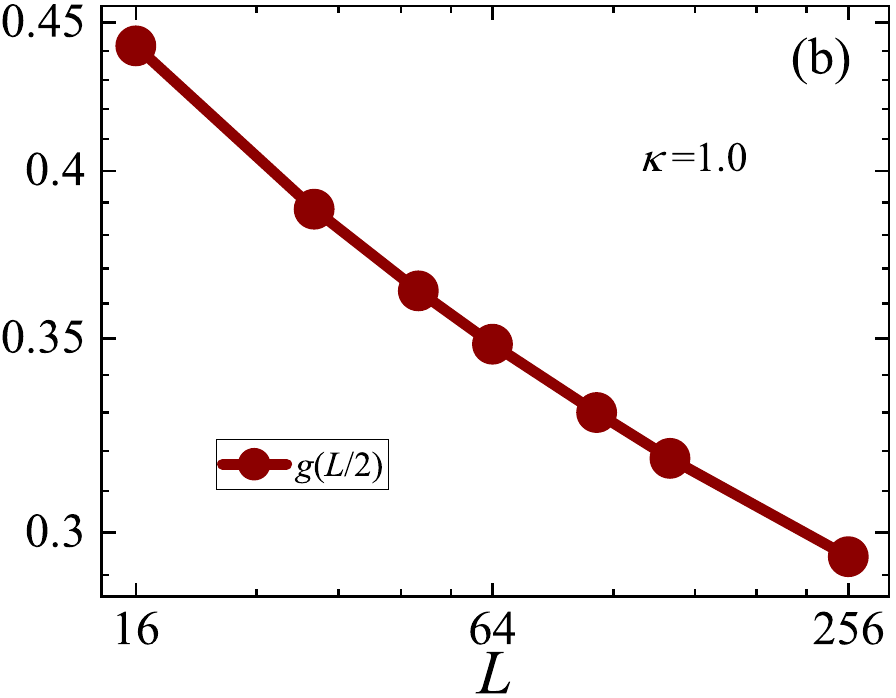}
\caption{(a) Log-log plot of $g(L/2)$ and the scaled magnetic fluctuations $\chi_0 L^{-2}$, $\chi_1 L^{-2}$ and $\chi_2 L^{-2}$ versus $L$ at the special transition point $\kappa_s=0.622\,2$. The constants $a_0$, $a_1$ and $a_2$ are used for data collapse. The dashed line has the slope $-0.650$ and represents the FSS behavior $g(L/2) \sim L^{-\eta}$ and $\chi_{{\bf k}} \sim L^{2y_h-2}$ with $\eta=0.650$ and $y_h=1.675$. (b) Log-log plot of $g(L/2)$  versus $L$ for the extraordinary transition at $\kappa=1$. In both panels, statistical errors are much smaller than the size of symbols.}~\label{SM_fig45}
\end{figure*}

As shown in Fig.~\ref{SM_fig3}, the $r$-dependent behavior converges to the power law $g(r) \sim r^{-\eta}$, with $\eta \approx 0.650$. From Fig.~\ref{SM_fig45}(a), it is confirmed that $g(L/2)$ decays as $g(L/2) \sim L^{-\eta}$ with $\eta \approx 0.650$, and  the magnetic fluctuations $\chi_0$, $\chi_1$ and $\chi_2$ {\it all} scale as $\chi_{{\bf k}}  \sim L^{2y_h-2}$ with $y_h \approx 1.675$. Hence, the magnetic fluctuations at zero and non-zero Fourier modes share the same leading scaling exponent. This feature is a consequence of the standard scenario for continuous transition, which is not applicable to the extraordinary-log transition. For a comparison, the size-dependent behavior of $g(L/2)$ for the extraordinary transition at $\kappa=1$ is demonstrated in Fig.~\ref{SM_fig45}(b).

\bibliography{papers2}

\begin{table*}
			\begin{center}
				\caption{Fits of $g(L/2)$ data to (\ref{fite1}).}
				\label{fit_result_1A}
				\begin{tabular}[t]{p{2cm}p{2cm}p{2cm}p{2cm}p{2cm}p{1.2cm}}
					\hline
					\hline
					$\kappa$                         &$L_{\rm min}$               & $\chi^2$/DF & $A$        & $l_0$          & $\hat{\eta}'$         \\
					\hline
					1.0 & 8      & 105.10/5                  & 0.853(2) & 0.844(6)   & 0.611(1)  \\
					& 16     & 2.91/4                    & 0.822(4) & 0.94(1)    & 0.596(2)  \\
					& 32     & 0.66/3                    & 0.813(7) & 0.97(2)    & 0.592(3)  \\
					& 48     & 0.58/2                    & 0.81(1)  & 0.98(4)    & 0.591(5)  \\
					& 64     & 0.58/1                    & 0.81(2)  & 0.98(6)    & 0.590(7)  \\
					
					1.5 & 8      & 4.39/5                    & 1.385(5) & 0.206(3)   & 0.597(1)  \\
					& 16     & 2.63/4                    & 1.395(9) & 0.201(5)   & 0.599(2)  \\
					& 32     & 2.41/3                    & 1.39(2)  & 0.205(10)  & 0.598(5)  \\
					& 48     & 0.04/2                    & 1.36(3)  & 0.22(2)    & 0.590(7)  \\
					& 64     & 0.03/1                    & 1.35(4)  & 0.22(2)    & 0.59(1)   \\
					
					3.0 & 8      & 3.39/5                    & 2.29(2)  & 0.0136(5)  & 0.582(2)  \\
					& 16     & 3.39/4                    & 2.29(3)  & 0.0136(7)  & 0.582(4)  \\
					& 32     & 2.33/3                    & 2.25(5)  & 0.015(1)   & 0.576(7)  \\
					& 48     & 2.23/2                    & 2.23(7)  & 0.015(2)   & 0.574(10) \\
					& 64     & 1.97/1                    & 2.20(10) & 0.017(4)   & 0.57(1)   \\
				
					5.0 & 8      & 4.79/5                    & 3.09(4)  & 0.00056(4) & 0.569(4)  \\
					& 16     & 3.10/4                    & 3.15(6)  & 0.00050(6) & 0.575(6)  \\
					& 32     & 1.69/3                    & 3.3(1)   & 0.00041(9) & 0.58(1)   \\
					& 48     & 0.49/2                    & 3.1(2)   & 0.0005(2)  & 0.57(1)   \\
					& 64     & 0.38/1                    & 3.2(2)   & 0.0005(2)  & 0.58(2)   \\
			
					\hline
					\hline
				\end{tabular}
			\end{center}
		\end{table*}

       \begin{table*}
			\begin{center}
            \caption{Fits of $g(L/2)$ data to $g(L/2)=AL^{-\eta}$.}
				\label{fit_result_1B}
				\begin{tabular}[t]{p{1.5cm}p{2cm}p{3cm}p{3cm}p{1.5cm}}
					\hline
					\hline
					$\kappa$                         &$L_{\rm min}$               & $\chi^2$/DF & $A$        & $\eta$	\\					
					\hline
					1.0
					& 32     & 29578.84/4                & 0.5979(2)   & 0.12938(6)    \\
					& 48     & 14146.05/3                & 0.5872(2)   & 0.12566(7)     \\
					& 64     & 6652.75/2                 & 0.5765(2)   & 0.12197(8)     \\
					& 96     & 977.67/1                  & 0.5607(3)   & 0.1165(1)      \\

					1.5
					& 32     & 16061.75/4                & 0.7361(1)   & 0.09833(4)      \\
					& 48     & 7517.73/3                 & 0.7274(2)   & 0.09584(5)      \\
					& 64     & 3228.71/2                 & 0.7180(2)   & 0.09323(7)     \\
					& 96     & 456.40/1                  & 0.7062(3)   & 0.08994(9)     \\

					3.0
					& 32     & 7436.35/4                 & 0.8724(1)   & 0.06614(3)      \\
					& 48     & 3449.23/3                 & 0.8676(1)   & 0.06500(4)       \\
					& 64     & 1680.08/2                 & 0.8631(2)   & 0.06393(4)     \\
					& 96     & 266.75/1                  & 0.8553(3)   & 0.06214(6)      \\

					5.0
					& 32     & 3190.77/4                 & 0.92996(10) & 0.04752(2)      \\
					& 48     & 1575.13/3                 & 0.9273(1)   & 0.04692(3)     \\
					& 64     & 684.47/2                  & 0.9243(2)   & 0.04626(4)      \\
					& 96     & 116.04/1                  & 0.9204(2)   & 0.04544(5)      \\

					\hline
					\hline					
				\end{tabular}
			\end{center}
		\end{table*}	

 \begin{table*}
			\begin{center}
            \caption{Fits of $g(L/2)$ data to $g(L/2)=AL^{-\eta}(1+BL^{-\omega})$.}
				\label{fit_result_1C}
				\begin{tabular}[t]{p{1.5cm}p{2cm}p{3cm}p{2.5cm}p{2.5cm}p{2.0cm}p{0.4cm}}
					\hline
					\hline
					$\kappa$                         &$L_{\rm min}$               & $\chi^2$/DF & $A$        & $\eta$	& $B$	& $\omega$ \\	
					\hline
					1.0
					& 8      & 4376.96/5                 & 0.5323(3)   & 0.10873(9)  & 1.859(5) & 1     \\
					& 16     & 1386.35/4                 & 0.5201(3)   & 0.1046(1)   & 2.208(8) & 1     \\
					& 32     & 278.84/3                  & 0.5061(5)   & 0.0998(2)   & 2.69(2)  & 1     \\
					& 48     & 61.44/2                   & 0.4987(7)   & 0.0973(3)   & 3.00(3)  & 1     \\
					& 64     & 15.47/1                   & 0.4942(10)  & 0.0958(3)   & 3.21(4)  & 1     \\
                    & 8      & 710.32/5                  & 0.3599(6)  & 0.0547(3)  & 1.722(6)  & 0.5 \\
					& 16     & 132.05/4                  & 0.3721(8)  & 0.0594(3)  & 1.593(8)  & 0.5 \\
					& 32     & 8.08/3                    & 0.382(1)   & 0.0632(5)  & 1.49(1)   & 0.5 \\
					& 48     & 1.01/2                    & 0.385(2)   & 0.0642(6)  & 1.46(2)   & 0.5 \\
					& 64     & 0.29/1                    & 0.386(2)   & 0.0647(8)  & 1.44(2)   & 0.5 \\				
					1.5
					& 8      & 4579.64/5                 & 0.6968(2)   & 0.08852(6)  & 0.837(2) & 1     \\
					& 16     & 1369.33/4                 & 0.6844(3)   & 0.08520(9)  & 1.063(5) & 1     \\
					& 32     & 285.77/3                  & 0.6700(5)   & 0.0814(1)   & 1.41(1)  & 1     \\
					& 48     & 46.02/2                   & 0.6616(8)   & 0.0793(2)   & 1.66(2)  & 1     \\
					& 64     & 6.55/1                    & 0.657(1)    & 0.0781(3)   & 1.82(3)  & 1     \\
                    & 8      & 31.15/5                   & 0.5867(5)  & 0.0641(1)  & 0.689(2)  & 0.5 \\
					& 16     & 31.06/4                   & 0.5868(7)  & 0.0641(2)  & 0.688(3)  & 0.5 \\
					& 32     & 13.76/3                   & 0.583(1)   & 0.0631(3)  & 0.711(7)  & 0.5 \\
					& 48     & 1.22/2                    & 0.579(2)   & 0.0621(4)  & 0.737(10) & 0.5 \\
					& 64     & 0.09/1                    & 0.577(2)   & 0.0616(6)  & 0.75(2)   & 0.5 \\				
					3.0
					& 8      & 2667.58/5                 & 0.8527(2)   & 0.06205(4)  & 0.343(2) & 1     \\
					& 16     & 868.60/4                  & 0.8450(3)   & 0.06037(6)  & 0.457(3) & 1     \\
					& 32     & 183.63/3                  & 0.8355(4)   & 0.05836(10) & 0.638(8) & 1     \\
					& 48     & 47.01/2                   & 0.8299(7)   & 0.0572(1)   & 0.77(1)  & 1     \\
					& 64     & 12.95/1                   & 0.8262(9)   & 0.0565(2)   & 0.86(2)  & 1     \\
                    & 8      & 274.79/5                  & 0.7990(4)  & 0.05275(8) & 0.256(1)  & 0.5 \\
					& 16     & 109.79/4                  & 0.7937(6)  & 0.0517(1)  & 0.275(2)  & 0.5 \\
					& 32     & 24.53/3                   & 0.7862(10) & 0.0503(2)  & 0.305(4)  & 0.5 \\
					& 48     & 9.18/2                    & 0.782(1)   & 0.0495(3)  & 0.322(6)  & 0.5 \\
					& 64     & 4.43/1                    & 0.779(2)   & 0.0490(4)  & 0.336(9)  & 0.5 \\					
					5.0
					& 8      & 1279.39/5                 & 0.9196(2)   & 0.04551(3)  & 0.168(1) & 1     \\
					& 16     & 396.88/4                  & 0.9150(2)   & 0.04458(5)  & 0.230(2) & 1     \\
					& 32     & 100.69/3                  & 0.9097(4)   & 0.04355(8)  & 0.321(6) & 1     \\
					& 48     & 15.82/2                   & 0.9060(5)   & 0.0428(1)   & 0.40(1)  & 1     \\
					& 64     & 4.97/1                    & 0.9042(8)   & 0.0425(1)   & 0.44(2)  & 1     \\
                    & 8      & 178.60/5                  & 0.8910(3)  & 0.04098(6) & 0.1227(9) & 0.5 \\
					& 16     & 60.06/4                   & 0.8871(5)  & 0.04030(9) & 0.135(1)  & 0.5 \\
					& 32     & 20.58/3                   & 0.8828(8)  & 0.0396(1)  & 0.150(3)  & 0.5 \\
					& 48     & 2.62/2                    & 0.879(1)   & 0.0390(2)  & 0.163(4)  & 0.5 \\
					& 64     & 1.54/1                    & 0.878(2)   & 0.0387(3)  & 0.169(7)  & 0.5 \\						
					\hline
					\hline					
				\end{tabular}
			\end{center}
		\end{table*}

		\begin{table*}
			\begin{center}
				\caption{Fits of $\chi_0$ to (\ref{fite2}).}
				\label{fit_result_2}
				\begin{tabular}[t]{p{2cm}p{2cm}p{2cm}p{2cm}p{2cm}p{1.2cm}}
					\hline
					\hline
					$\kappa$                         &$L_{\rm min}$               & $\chi^2$/DF & $A$          & $l_0$          & $\hat{\eta}'$     \\
					\hline					
					1.0
					& 8      & 785.73/5                  & 0.936(2) & 0.814(5)   & 0.6495(9) \\
					& 16     & 58.85/4                   & 0.867(3) & 1.008(9)   & 0.619(1)  \\
					& 32     & 3.46/3                    & 0.833(5) & 1.13(2)    & 0.603(2)  \\
					& 48     & 0.08/2                    & 0.823(7) & 1.18(3)    & 0.598(4)  \\
					& 64     & 0.02/1                    & 0.82(1)  & 1.19(5)    & 0.597(5)  \\

					1.5
					& 8      & 72.23/5                   & 1.481(4) & 0.203(2)   & 0.622(1)  \\
					& 16     & 13.06/4                   & 1.436(7) & 0.225(4)   & 0.611(2)  \\
					& 32     & 4.39/3                    & 1.40(1)  & 0.246(8)   & 0.602(3)  \\
					& 48     & 0.56/2                    & 1.37(2)  & 0.27(1)    & 0.594(5)  \\
					& 64     & 0.35/1                    & 1.36(3)  & 0.27(2)    & 0.592(7)  \\

					3.0
					& 8      & 21.51/5                   & 2.45(2)  & 0.0121(3)  & 0.604(2)  \\
					& 16     & 9.13/4                    & 2.39(2)  & 0.0135(6)  & 0.596(3)  \\
					& 32     & 3.39/3                    & 2.32(4)  & 0.016(1)   & 0.585(5)  \\
					& 48     & 1.15/2                    & 2.26(5)  & 0.018(2)   & 0.577(7)  \\
					& 64     & 0.92/1                    & 2.23(7)  & 0.019(3)   & 0.57(1)   \\

					5.0
					& 8      & 5.20/5                    & 3.30(3)  & 0.00046(3) & 0.588(3)  \\
					& 16     & 1.96/4                    & 3.23(5)  & 0.00052(5) & 0.582(4)  \\
					& 32     & 1.57/3                    & 3.27(9)  & 0.00048(8) & 0.586(8)  \\
					& 48     & 0.12/2                    & 3.2(1)   & 0.0006(1)  & 0.58(1)   \\
					& 64     & 0.02/1                    & 3.2(2)   & 0.0005(2)  & 0.58(2)   \\

					\hline
					\hline
				\end{tabular}
			\end{center}
		\end{table*}

\begin{table*}	
			\begin{center}
			\caption{Fits of $\chi_1$ and $\chi_2$ to (\ref{fite3}).}
			\label{fit_result_3a}
				\begin{tabular}[t]{p{2cm}p{2cm}p{2cm}p{2cm}p{2cm}p{2cm}p{1.cm}}
					\hline
					\hline
					Quantity   &$\kappa$      &$L_{\rm min}$   & $\chi^2$/DF & $A$  & $l_0$  & $\hat{\eta}$ \\
					\hline					
					$\chi_1$
					&  1.0   & 48 & 3.79/2   & 0.35(7)       & 0.33(6)    & 2.14(7)  \\
					&     & 64 & 0.18/1   & 0.24(6)       & 0.5(1)     & 2.01(9)  \\
					&     & 96 & 0.00/0   & 0.2(1)        & 0.6(4)     & 1.9(2)   \\
					&   1.5    & 48 & 0.31/2   & 0.3(1)        & 0.11(5)    & 1.9(1)   \\
					&     & 64 & 0.03/1   & 0.3(1)        & 0.14(9)    & 1.8(2)   \\
					&     & 96 & 0.00/0   & 0.2(3)        & 0.2(3)     & 1.8(4)   \\
					&  3.0    & 32 & 2.44/3   & 1.1(9)        & 0.002(2)   & 2.1(2)   \\
					&     & 48 & 1.60/2   & 3.1(56)       & 0.0003(8)  & 2.4(5)   \\
					&     & 64 & 1.04/1   & 1.0(18)       & 0.002(5)   & 2.0(5)   \\
					&   5.0    & 32 & 3.14/3   & 1.2(23)       & 0.0000(2)  & 2.0(5)   \\
					&     & 48 & 1.26/2   & 0.2(2)        & 0.003(8)   & 1.4(4)   \\
					&     & 64 & 1.22/1   & 0.2(5)        & 0.001(7)   & 1.5(7)   \\
					\hline
					$\chi_2$
					&  1.0   & 48 & 0.12/2   & 0.32(7)       & 0.27(5)    & 2.36(8)  \\
					&     & 64 & 0.00/1   & 0.29(9)       & 0.29(8)    & 2.3(1)   \\
					&     & 96 & 0.00/0   & 0.3(2)        & 0.3(2)     & 2.3(3)   \\
					&  1.5   & 48 & 0.29/2   & 0.3(1)        & 0.07(3)    & 2.1(1)   \\
					&     & 64 & 0.08/1   & 0.3(2)        & 0.09(6)    & 2.1(2)   \\
					&     & 96 & 0.00/0   & 0.2(2)        & 0.1(2)     & 1.9(5)   \\
					&   3.0  & 48 & 0.55/2   & 3.8(77)       & 0.0001(4)  & 2.6(6)   \\
					&     & 64 & 0.48/1   & 2.3(59)       & 0.000(1)   & 2.5(7)   \\
					&     & 96 & 0.00/0   & 0.1(4)        & 0.02(10)   & 1.7(9)   \\
					&   5.0   & 48 & 3.28/2   & 0.6(18)       & 0.0000(2)  & 2.0(8)   \\
					&     & 64 & 0.15/1   & 0.04(4)       & 0.02(7)    & 1.1(4)   \\
					&     & 96 & 0.00/0   & 0.2(14)       & 0.001(8)   & 1.6(20)  \\
					\hline
					\hline
				\end{tabular}
			\end{center}
	    	\end{table*}	
	    	
		\begin{table*}
			\begin{center}
				\caption{Fits of $\chi_1$ and $\chi_2$ to (\ref{fite3}), with the parameter $l_0$ being fixed at those of preferred fits in Table~\ref{fit_result_2}.}
			\label{fit_result_3b}
				\begin{tabular}[t]{p{2cm}p{2cm}p{2cm}p{2cm}p{2cm}p{2cm}p{1.2cm}}
					\hline
					\hline
					Quantity   &$\kappa$      &$L_{\rm min}$   & $\chi^2$/DF & $A$  & $l_0$  & $\hat{\eta}$ \\
					\hline
					$\chi_1$	
					&   1.0   & 48  & 72.00/3    & 0.1023(4)  & 1.13    & 1.676(3)   \\
					&     & 64  & 16.07/2    & 0.1050(6)  & 1.13    & 1.693(4)   \\
					&     & 96  & 1.10/1     & 0.1078(9)  & 1.13    & 1.709(6)   \\
					&     & 128 & 0.00/0     & 0.109(1)   & 1.13    & 1.716(8)   \\		
					&   1.5   & 48  & 4.66/3     & 0.174(1)   & 0.246   & 1.687(4)   \\
					&     & 64  & 0.91/2     & 0.176(2)   & 0.246   & 1.695(6)   \\
					&     & 96  & 0.03/1     & 0.178(3)   & 0.246   & 1.702(9)   \\
					&     & 128 & 0.00/0     & 0.179(4)   & 0.246   & 1.70(1)    \\	
					&   3.0   & 48  & 6.21/3     & 0.257(4)   & 0.016   & 1.635(7)   \\
					&     & 64  & 1.90/2     & 0.264(5)   & 0.016   & 1.647(10)  \\
					&     & 96  & 1.66/1     & 0.267(9)   & 0.016   & 1.65(2)    \\
					&     & 128 & 0.00/0     & 0.28(1)    & 0.016   & 1.67(2)    \\
					&   5.0   & 48  & 1.53/3     & 0.38(1)    & 0.00046 & 1.65(1)    \\
					&     & 64  & 1.25/2     & 0.38(2)    & 0.00046 & 1.65(2)    \\
					&     & 96  & 0.90/1     & 0.37(2)    & 0.00046 & 1.64(2)    \\
					&     & 128 & 0.00/0     & 0.39(3)    & 0.00046 & 1.66(3)    \\
					\hline
					$\chi_2$				
					&  1.0   & 48  & 102.22/3   & 0.0684(3)  & 1.13    & 1.779(3)   \\
					&     & 64  & 41.47/2    & 0.0702(4)  & 1.13    & 1.796(3)   \\
					&     & 96  & 5.50/1     & 0.0731(6)  & 1.13    & 1.822(6)   \\
					&     & 128 & 0.00/0     & 0.0747(9)  & 1.13    & 1.835(8)   \\		
					&   1.5    & 48  & 10.47/3    & 0.1099(9)  & 0.246   & 1.762(4)   \\
					&     & 64  & 3.05/2     & 0.112(1)   & 0.246   & 1.773(6)   \\
					&     & 96  & 0.13/1     & 0.115(2)   & 0.246   & 1.785(9)   \\
					&     & 128 & 0.00/0     & 0.115(3)   & 0.246   & 1.79(1)    \\
					&  3.0    & 48  & 7.09/3     & 0.156(2)   & 0.016   & 1.693(7)   \\
					&     & 64  & 3.00/2     & 0.160(3)   & 0.016   & 1.705(9)   \\
					&     & 96  & 0.00/1     & 0.167(5)   & 0.016   & 1.73(1)    \\
					&     & 128 & 0.00/0     & 0.167(8)   & 0.016   & 1.72(2)    \\	
					&    5.0  & 48  & 3.53/3     & 0.206(6)   & 0.00046 & 1.66(1)    \\
					&     & 64  & 1.13/2     & 0.214(9)   & 0.00046 & 1.68(2)    \\
					&     & 96  & 0.00/1     & 0.20(1)    & 0.00046 & 1.66(2)    \\
					&     & 128 & 0.00/0     & 0.20(2)    & 0.00046 & 1.66(3)    \\
					\hline
					\hline					
				\end{tabular}
			\end{center}
		\end{table*}

				\begin{table*}
			\begin{center}
				\caption{Fits of $\Upsilon$ to (\ref{fite4}). In the last column, ``-'' means that the correction term $BL^{-1}$ is not included.}
				\label{fit_result_4}
				\begin{tabular}[t]{p{2cm}p{2cm}p{2cm}p{2cm}p{2cm}p{1.1cm}}
					\hline
					\hline
					$\kappa$    &$L_{\rm min}$    & $\chi^2$/DF   & $\alpha$   & $A$   & $B$ \\
					\hline
					1.0
					& 8      & 5.46/4     & 0.255(3)  & 0.41(2)  & 1.60(8)   \\
					& 16     & 3.33/3     & 0.265(7)  & 0.32(6)  & 2.2(4)    \\
					& 32     & 2.51/2     & 0.25(2)   & 0.4(2)   & 0.8(15)   \\
					& 48     & 0.73/1     & 0.29(3)   & 0.1(3)   & 6.4(45)   \\
					& 32     & 2.80/3     & 0.245(4)  & 0.52(3)  & -         \\
					& 48     & 2.79/2     & 0.245(6)  & 0.52(5)  & -         \\
					& 64     & 1.64/1     & 0.252(9)  & 0.46(8)  & -         \\
					1.5
					& 8      & 2.52/5     & 0.270(2)  & 1.08(2)  & 0.82(6)   \\
					& 16     & 2.47/4     & 0.271(7)  & 1.07(6)  & 0.9(4)    \\
			        & 32     & 4.23/4     & 0.261(3)  & 1.16(2)  & -         \\
					& 48     & 0.99/3     & 0.270(6)  & 1.09(5)  & -         \\
					& 64     & 0.23/2     & 0.276(9)  & 1.04(8)  & -         \\
					& 96     & 0.10/1     & 0.27(2)   & 1.1(2)   & -         \\
					3.0
					& 8      & 8.21/5     & 0.278(2)  & 2.59(2)  & 0.45(7)   \\
					& 16     & 7.40/4     & 0.283(6)  & 2.54(6)  & 0.8(4)    \\
					& 32     & 7.39/3     & 0.28(2)   & 2.5(1)   & 0.9(15)   \\
					& 48     & 7.38/2     & 0.28(3)   & 2.6(3)   & 0.3(44)   \\	
					& 16     & 11.95/5    & 0.270(1)  & 2.662(8) & -         \\
					& 32     & 7.72/4     & 0.276(3)  & 2.62(2)  & -         \\
					& 48     & 7.38/3     & 0.278(5)  & 2.59(4)  & -         \\
					& 64     & 7.14/2     & 0.275(8)  & 2.62(7)  & -         \\
					5.0
					& 8      & 2.23/5     & 0.276(2)  & 4.49(2)  & 0.19(7)   \\
					& 16     & 1.73/4     & 0.281(6)  & 4.45(6)  & 0.4(4)    \\
					& 32     & 1.17/3     & 0.27(2)   & 4.6(2)   & -0.7(15)  \\
					& 8      & 9.77/6     & 0.2701(6) & 4.538(3) & -         \\
					& 16     & 3.19/5     & 0.273(1)  & 4.520(8) & -        \\
					& 32     & 1.37/4     & 0.277(3)  & 4.49(2)  & -         \\
					& 48     & 1.32/3     & 0.276(5)  & 4.50(4)  & -         \\
					& 64     & 0.76/2     & 0.271(9)  & 4.54(8)  & -         \\		
					\hline
					\hline
				\end{tabular}
			\end{center}
		\end{table*}